# Origin of the thickness-dependent low-temperature enhancement of spin Seebeck effect in YIG films

E. J. Guo, A. Kehlberger, J. Cramer, G. Jakob, and M. Kläui [a]

*Institut für Physik, Johannes Gutenberg-Universität Mainz, 55099 Mainz, Germany*

**Abstract**

The temperature dependent longitudinal spin Seebeck effect (SSE) in heavy metal (HM)/$Y_3Fe_5O_{12}$ (YIG) bilayers is investigated as a function of different magnetic field strength, different HM detection material, and YIG thickness ranging from nm to mm. A large enhancement of the SSE signal is observed at low temperatures leading to a peak of the signal amplitude. We demonstrate that this enhancement shows a clear dependence on the film thickness, being more pronounced for thicker films and vanishing for films thinner than 600 nm. The peak temperature depends on the applied magnetic field strength as well as on the detection material and interface, revealing a more complex behavior beyond the currently discussed phonon-magnon coupling mechanism that considers only bulk effects. While the thickness dependence and magnetic field dependence can be well explained in the framework of the magnon-driven SSE by taking into account the frequency dependent propagation length of thermally excited magnons in the bulk material, the temperature dependence of the SSE is significantly influenced by the interface coupling to an adjacent detection layer. This indicates that previously neglected interface effects play a key role and that the spin current traversing the interface and being detected in the HM depends differently on the magnon frequency for different HMs.

---

[a] Author to whom correspondence should be addressed. Electronic mail: klaeui@uni-mainz.de

The spin Seebeck effect (SSE) describes the phenomenon that spin currents can be thermally excited by a temperature gradient across magnetic materials [1-3]. These spin currents can be injected into an adjacent heavy metal layer with a high spin–orbit coupling, which allows one to electrically detect the spin currents due to the inverse spin Hall effect (ISHE) [1-9]. The SSE has been observed not only in metals and semiconductors, but even in magnetic insulators. Thus this effect provides a suitable platform to investigate the origin of the spin current that crosses the interfaces [8, 9]. In a magnetic insulator [3, 6-8], the excitations of the localized spin-polarized electrons are the primary source of the spin currents due to the absence of itinerant electrons. The injected thermal energy can only propagate through quasi-particle excitations, of either the magnetization (magnons) or the real space atom displacements (phonons). It is crucial to have a better understanding of relevant interactions between magnons and phonons and the corresponding length scales that govern magnon transport. The first investigation of the phonon-magnon interaction in magnetic insulators was conducted by Adachi *et al.*, reporting a giant enhancement of SSE in $LaY_2Fe_5O_{12}$ at low temperatures [11]. This observation challenged the magnon-mediated SSE theory that was initially developed to explain the room-temperature results [12], but fails to interpret the low-temperature enhancement of the SSE. To explain the low temperature enhancement, Adachi *et al.* proposed the phonon-drag SSE scenario based on a theoretical model [11, 13]. In this phenomenological model, the temperature dependence of the phonon life time is involved, which reaches a maximum at low temperatures. This leads to a peak of the thermal conductance at similar temperatures as the peak observed for the SSE. Based on this, they propose a strong interaction between the phonons and magnons, which are responsible for the heat transport in the system. The phonons flow along the thermal gradient and interact with thermally excited magnons. These phonons "drag" the magnons, thereby enhancing the pumped spin current into the adjacent Pt detection layer. Thus, the phonon-magnon coupling is suggested to explain the observed enhancement of SSE signal at low temperature [11, 13]. However, on one hand, the observed transport of magnons over a long distance of up to millimeters in magnetic insulators implies a relative weak interaction with phonons and impurities [3]. On the other hand, contradicting the phonon-magnon drag model, recent measurements of the temperature dependent thermal conductance of YIG single crystals show that the phonon contribution to the thermal conductivity reaches its maximum at around 25 K [15], which is ~50 K lower than the observed peak in the SSE [11, 13]. Additionally the magnon contribution to the thermal conductance has been found to be less than 15% at 2 K and decreases rapidly with increasing temperatures, suggesting an even weaker coupling between the magnons and phonons at high temperatures [15]. Therefore, the validity of the phonon-drag SSE mechanism in explaining the temperature dependence of the SSE in YIG is not obvious and the genuine origin of low temperature enhancement of SSE needs to be clarified. In addition to these discrepancies, the SSE signals are indirectly probed via the ISHE so that not only the bulk properties of magnetic materials can be important but the injection efficiency of a thermal spin current across the interface, described by the spin mixing conductance, is also crucial for the detected SSE voltages [16-18]. Previous works have shown that the SSE voltage is enhanced dramatically when atomically smooth and well-crystalline interfaces are introduced [16], and different ISHE detection materials which possess different spin Hall angles also lead to different SSE signals [6]. Thus far, there is no information about the temperature dependent spin current transparency of the interfaces and its implication for the low temperature enhancement of the SSE signals, which needs to be ascertained to fully understand the SSE signals and their temperature dependence.

In this paper, we study the temperature dependence of the SSE in a YIG single crystal and thin film samples spanning a wide thickness range to reveal the bulk and interface origins leading to complex behavior. Our results reveal that the SSE peaks at specific temperatures strongly dependent on the film thickness, showing a shifting towards higher temperatures as the film thickness decreases. The peak temperature can be additionally altered by the magnetic field, which causes a shift towards lower temperatures for higher magnetic fields. To understand these findings, we introduce a simple phenomenological model, capable of explaining the observed thickness and magnetic field dependence by a solely magnonic propagation. Beyond this simple model we find a strong dependence of the peak temperature on the ISHE detection material which reveals a more complex origin of the SSE signals that depend on bulk thermal spin current properties as well as on the detection layer material and interface.

The samples used in the present study are YIG thin films grown on $Gd_3Ga_5O_{12}$ (GGG) substrate (5×10×0.5 mm³) by liquid phase epitaxy (LPE) and a (111)-oriented YIG slab (5×10×1 mm³). The thickness of YIG thin films ranges from 150 nm to 50 μm. A thin HM layer is sputtered on the top of the YIG samples with a thickness of 5.5 nm. We further pattern the HM layer into strips by optical lithography and argon ion beam etching, for comparable conditions between all of our samples. The HM strips have a length of $L_{Pt}$ = 4 mm and a width of w=100 μm. As indicated in the inset of Fig. 1(c), our SSE measurements are carried out in the longitudinal configuration with the temperature gradient and thermally excited spin currents applied in the out-of-plane direction [7]. The samples are sandwiched between a resistive heater and a thermal sensor, which are mounted onto a copper heat sink. To avoid electric short circuits but to allow for a good heat flow in the out-of-plane direction, all thermo-elements are structured on $Al_2O_3$ substrates and uniformly glued together with a high thermally conducting grease. The temperature dependent measurements are performed in a cryostat with variable temperature insert in the temperature range from 30 to 300 K. The thermal gradient can be varied by changing the heat currents of the resistive heater. We start with Pt as the HM layer and an advantage of our experimental setup is that the Pt strips on the top and ground surfaces of the sample can be utilized as resistance-temperature sensors to determine the temperature differences (Δ*T*) across the Pt/YIG bilayer. The temperature used for the SSE measurements is the temperature obtained from the Pt strip on the YIG surface, providing the best possible determination of the actual system temperature. The in-plane magnetic field *H* is applied along the *y*-axis, perpendicular oriented to the long axis of the Pt strips (*x*-axis) in order to provide highest signals due to the ISHE, and the applied field is strong enough to ensure full alignment of the magnetization along this axis for YIG. The SSE signal is detected electrically by measuring the voltage drop at the two ends of the Pt strip (along *x*-axis). Our experiments are carried out in the temperature sweep mode, in which the magnetic field is kept constant while the temperature is swept twice (firstly at +*H*, then at –*H*) from 300 K down to 30 K at a rate of -0.25 K/min. The low sweep rate provides sufficient time for the system to stabilize to the thermal equilibrium between the sample and the cryostat during the temperature drop, yielding reproducible conditions for the temperature sweep measurements. The SSE signals are determined as $V_{SSE} = [V_{SSE}(+H)-V_{SSE}(-H)]/2$. In order to quantitatively compare the SSE signals for different samples, the SSE coefficients are calculated by $\sigma_{SSE} = (V_{SSE}/\Delta T) \times (L_z/L_{Pt})$, where the $L_z$ is the total thickness of the YIG slab or YIG/GGG samples.

The $\sigma_{SSE}$ as a function of temperature for a YIG slab is shown in Figure 1(a). The SSE signal gradually increases with decreasing temperature and shows a pronounced peak at around 75 K which is qualitatively consistent with previous reports [11, 13]. We also conduct the SSE measurements by sweeping the magnetic field at fixed temperatures, which are commonly used in the previous reports [5-13]. The SSE voltages flip sign when the direction of the magnetic field reverses and we extract the $V_{SSE}$ from the difference of the saturated signal as indicated by the open symbols in the Fig. 1(a). The good agreement between both measurement methods validates the reproducibility and reliability of the temperature sweep mode results which we use for all further measurements in this study. Figure 1(b) presents the temperature dependent $\sigma_{SSE}$ for YIG thin films with thicknesses ranging from 150 nm to 50 $\mu$m. The low temperature enhancement of the SSE is observed for thicker YIG films and is reduced as the film thickness decreases. There is no visible peak for films with a thickness below 600 nm. The SSE peak shifts to higher temperatures and its width increases as the film thickness goes down as shown in Fig. 1(c). To exclude any possible artifacts involved in the SSE measurements, additional effects which exhibit a temperature dependence and may influence the peak position of the SSE signal are analyzed: the temperature dependent resistances of Pt strips are monitored simultaneously during the SSE measurements. The Pt resistances decrease monotonically as temperature decreases. We calculate the $I_{SSE}$ (= $V_{SSE} / R_{Pt}$), which yields a shift of the peak position of ~5 K, which is identical for all measured samples due to the similar quality of the Pt stripes. Other temperature dependent effects might result from the spin Hall angle of the Pt and from the spin mixing conductance across the Pt/YIG interfaces. From literature we find that those show a negligible variation with temperature for the investigated temperature regime [19]. Therefore, the only possible mechanisms responsible for the strong temperature dependence of the SSE are the bulk effects in the YIG material and the interface effects with the adjacent detection HM layers.

In the YIG bulk materials, the phonons coexist with the magnons and contribute to the thermal conductance at all temperatures. Our recent measurements of the thermal conductance as a function of temperature on the identical YIG samples reveal a minor shift of the peak position of the thermal conductance with decreasing film thickness [20]. The absence of a pronounced thickness dependence of the thermal conductivity peak in combination with the mismatch between both peak hints towards a less important contribution of the magnon-phonon coupling than previously assumed. So, a possibly more apt approach to explain the effect is to consider the concept of a magnonic flux solely, based on the idea of a temperature dependent thermal magnon propagation length [8, 23, 24]. In our previous work [8], we demonstrated that a finite magnon propagation length can qualitatively explain the observed increase and saturation of the $V_{SSE}$ in YIG thin films with increasing film thickness. This behavior is well described by an atomistic spin model which is capable to describe thermally excited magnon accumulation generated by the temperature gradient using [8, 21],

$$V_{SSE} \propto 1-\exp(-L/\xi) \tag{1}$$

where $\xi$ is the characteristic length of the SSE, regarding it as mean propagation length of all thermally excited magnons over the whole frequency spectrum. To determine the magnon propagation length, we plot $\sigma_{SSE}$ as a function of film thickness at temperatures of 250 K, 120 K, and 50 K, respectively, as shown in the Figs. 2(a)-(c). $\sigma_{SSE}$ is strongly dependent on the film thickness and follows the same trend as in our previous results [8]: $\sigma_{SSE}$ increases gradually with increasing thickness and saturates above a critical thickness. The SSE saturation shows that the magnons possess a finite propagation length. This means that only magnons excited at distances smaller than $\xi$ from the interface contribute to the spin currents which can be detected in the adjacent metal layer due to the ISHE. By fitting our results in Figs. 2(a)-(c) with Eq. (1), we can estimate the $\xi$ across the whole temperature range [Fig. 2(d)]. We find a monotonous increase of the magnon propagation length with decreasing temperature, following a functional dependency of $\xi \sim T^{-1}$. In the framework of their theoretical model, Ritzmann *et al.* have shown that $\xi$ is inverse propagation to Gilbert damping constant $\alpha$ ($\xi \sim \alpha^{-1}$) [18, 21], which leads to a longer propagation length in systems with lower magnetic damping. Additionally it is known from Lecraw and Spencer *et al.* [22] that the magnetic damping constant $\alpha$ of YIG crystal depends approximately linearly on the temperature (with slight variations due to the inherent micro-structure defects). This linear temperature dependence of the Gilbert damping in combination with its inverse proportionality leads to the observed ξ~ $T^{-1}$ dependence, which well agrees with our experiments [Fig. 2(d)]. Having understood the origin of the temperature dependence of the magnon propagation length, we can use this concept to explain the temperature dependence of the SSE signals. For thick YIG films, the efficient magnon propagation length increases with decreasing temperature, so more magnons can reach the detection layer, leading to an enhancement of SSE signal. In parallel the total amount of thermally excited magnons decreases as a function of temperature, which can be best approximated using the specific heat of the magnetic lattice. At a certain temperature, the increase of the propagation length is limited by the boundary scattering mechanism, identical to the analogous concept for phonons. Around this temperature, $\sigma_{SSE}$ reaches a maximum since the total amount of magnons, which can reach the detection layer cannot increase further. At even lower temperature, the magnon propagation length stays constant, while the amount of thermally excited magnons decreases leading to a decrease of $\sigma_{SSE}$. Since the propagation length is limited in thinner YIG films at higher temperatures, the peak temperature is shifted towards higher temperatures as the film thickness decreases and the overall amplitude of SSE signal decreases as well. For films showing an absence of a peak, the mean magnon propagation length is of the order of the film thickness at room temperature, as shown in Fig. 2(d), leading to a temperature dependence of the $\sigma_{SSE}$ dominated by the specific heat of the magnetic lattice. Since the excitation of thermal magnons follows a Bose-Einstein distribution it is smeared out for higher temperatures, leading to a broadening of the peak. This combination of a temperature dependent magnon propagation length and the total amount of thermal magnons can phenomenologically explain the presence of the peak in the SSE signal.

In our experiments, the ISHE is used to detect the magnonic spin current, and so the magnon propagation length that we obtain represents only a mean value of all thermally excited magnons that are able to reach the detection layer, independent of their energy. Applying a high magnetic field can suppress low-energy magnonic excitations and thus provide a possible way to manipulate the magnonic distribution in magnetic insulators [15, 30, 31]. In our recent work [23], we measured the magnetic field suppression of the SSE signal for a YIG single crystal and YIG films with various thicknesses at room temperature. For YIG single crystals, the field suppression reaches ~40% at 9 T, while this value drops to less than 0.1% for 300 nm-thick YIG thin films under the same conditions. In this work, in order to rule out any possible proximity effects [9, 24], the temperature dependent SSE measurements are carried out using a Pt/Cu/YIG(50 μm) sample by inserting an ultra-thin Cu interlayer with thickness of 2 nm between the Pt and YIG. Figure 3(a) shows the field suppression of the SSE for the Pt/Cu/YIG(50 μm) sample at various temperatures. The SSE field suppression ratio $\delta V_{SSE}$ {=[1-$V_{SSE}$(9T)/$V_{SSE}$(0.2T)]×100%} drops monotonously from ~30% at room temperature to ~9% at 30 K, as illustrated in Fig. 3(b). The temperature dependent SSE under variable magnetic fields from 0.2 T to 9 T is shown in Fig. 3(c). We find a significant reduction of $\sigma_{SSE}$ under the magnetic field at high temperatures and the reduction becomes small with decreasing temperature, in agreement with the field sweep measurements shown in Fig. 3(b). This causes also a shift of the SSE peak of ~10 K towards lower temperatures. This effect of the field suppression cannot be explained by the opening of a magnon energy gap, since our experiments are carried out at relatively high temperatures. For YIG, the estimated magnon gap temperature ($\mu_B g_L H/k_B$, where $\mu_B$ is the Bohr magneton, $g_L$ is the Landé factor, and $k_B$ is the Boltzmann constant) at 9 T magnetic field is ~12 K by using $g_L$ = 2.046 for YIG, highlighting that the effect should be negligible at room temperature and become gradually more important as temperature decreases. However, while the total number of magnons at room temperature barely changes, when applying a high magnetic field, the low-frequency (energy) magnon spectrum will be cut off significantly [23]. This leads to a field suppression of the detected SSE signal at all temperatures when the low-frequency magnons are important. Our results, shown in Fig. 3(b), yield that the field suppression is reduced as the temperature is decreased down to our lowest probing temperature of 30 K. Based on the concept of a reduction of the effective magnon propagation length by the applied magnetic field [23], we can interpret our results as a reduction of the mean magnonic propagation length by the applied magnetic field, leading to a decreasing signal for all temperatures. As the magnonic propagation length increases for lower temperatures, it counteracts the reduced propagation length by the field suppression. With decreasing temperature, the propagation length increases until it becomes limited by the intrinsic boundary scattering mechanism limiting the magnon propagation, which leads to the observed shift of the SSE peak temperature to lower temperatures. Our interpretation is further strengthened by our observation that the peak temperature is independent of the applied heating power as it stays constant at ~ 82 K, as shown in Fig. 3(d). The magnitude of $V_{SSE}$ is confirmed to increase linearly with increasing $\Delta T$, since the temperature differences applied here are small enough so that the SSE stays in a linear response regime [25]. This reveals the validity of our magnon propagation model, which holds for all magnons of the whole excited frequency spectrum accessible by the applied temperature gradients.

Having determined the effects resulting from the bulk magnonic spin current, we next investigate effects that result from the previously neglected interface to the HM detector materials. So we concentrate on varying the interfaces to study the impact on the SSE in the HM/YIG hybrids. We know that the amplitudes of the room-temperature SSE signal vary strongly when changing the HM detection layers due to the different spin Hall angles that directly impact the detection efficiency [16-18]. In addition to the previously used Pt/YIG and Pt/Cu/YIG samples, we introduce Pd as a third spin detection layer for comparison. Thin Pd layers are sputtered on top of YIG surface with the same thickness as Pt layers. The three YIG films (50 μm thickness) are identical from the same wafer to exclude any variations from bulk YIG effects. Figure 4 shows a comparison of the temperature dependence of $\sigma_{SSE}$ in Pt/YIG, Pt/Cu/YIG and Pd/YIG samples. Compared to the Pt/YIG sample, the $\sigma_{SSE}$ of Pt/Cu/YIG sample is reduced owing to the partial loss of the injected spin currents during the spin diffusion across the Cu interlayer to the Pt layer [26]. The $\sigma_{SSE}$ of Pd/YIG is much smaller than that of Pt/YIG sample. This can be quantitatively understood due to the smaller spin Hall angle ($\theta_{SHE,Pd}$ =0.0064) and resistivity ($\rho_{Pd}$ = 0.25 μΩ m ) of Pd compared to Pt ($\theta_{SHE,Pt}$ = 0.013 ± 0.002, $\rho_{Pt}$ = 0.42 μΩ m) [27]. In order to compensate for these secondary effects (spin Hall angle, resistivity, *etc.*) that lead to different efficiencies for different detection materials, we normalize the $\sigma_{SSE}$ of three samples to their maximum values at the SSE peaks, respectively, as shown in the inset of Fig. 4. Surprisingly, the SSE peak position varies strongly from 100 K in Pt/YIG, to 66 K in Pt/Cu/YIG, and to 49 K in Pd/YIG, which cannot be explained by the different spin Hall angles or the spin absorption in the Cu. We find that by simply changing the interfaces in contact with the YIG films, the temperature dependent SSE is significantly changed. These unexpected results cannot be explained by the scenario of phonon-drag SSE due to the phonon-magnon coupling [10, 12]. If the phonon-drag spin distribution in the YIG bulk material dominates the SSE signal at low temperatures, the SSE peak position should not be modulated by altering the interface. Therefore, our results strongly suggest that the interface thus the spin mixing conductance also depends on the magnon frequency. For the three samples with identical YIG, the numbers of thermally excited magnons are equal under the same temperature gradient. Since the low-energy magnons dominate the SSE signal as shown above, the difference in the temperature dependence of the SSE clearly reflects the variation of the transparency of the low-energy magnons through the different interface conditions. As we know from the thickness dependent SSE results in Pt/YIG hybrids, a larger SSE signal and a lower SSE peak temperature are observed in thicker YIG films. The low-energy magnons' spin mixing conductances of the Pt/YIG interfaces stay nearly the same as all Pt layers are grown onto identical YIG films simultaneously in the same conditions [28, 29]. Therefore, it is clear that a lower SSE peak temperature indicates a larger amount of low-energy magnons that can reach the Pt/YIG interface and contribute to the detected ISHE voltages. Based on this analysis, we can semi-quantitatively conclude that the Pd/YIG interface possesses the highest transparency for the low-energy magnons, the Pt/Cu/YIG interface has an intermediate transparency, and the Pt/YIG interface shows the lowest efficiency of the low-energy magnons traversing into the HM layer. Thus the injection of the low-energy magnons varies depending on the adjacent metals, resulting in a dramatic change of temperature dependence of the SSE. These results reveal that the spin mixing conductance at the interface, as an important parameter for SSE, is magnon frequency dependent for different materials and dramatically influences the temperature dependence of the SSE, which was previously ignored.

In conclusion, we determine the temperature and thickness dependence of the SSE on YIG single crystals and thin films with different capping layers. The giant enhancement of the SSE at low temperatures is strongly dependent on the film thickness and this dependence is distinctly different from the thermal conductance measurements, indicating no dominating role of phonon-magnon coupling in this system. We can explain our results in the framework of pure thermally excited magnons in YIG based on the concept of a temperature and magnon-frequency dependent propagation length. The low-energy magnons dominate the SSE signals and can be suppressed under applied magnetic fields. We note that, during the preparation of this manuscript, Kikkawa *et al.* [30] and Jin *et al.* [31] have reported similar substantial field suppressions of the SSE signals, which are consistent with ours. In addition to these bulk spin current properties, we furthermore identify the interfaces as a major contributor to the temperature dependence of the SSE signals. A significant change in the SSE peak position is observed by changing different capping layers, demonstrating the spin current transmission through the interface to be also temperature and thus magnon-frequency dependent. Our results reveal that the temperature dependent SSE is not solely governed by the magnetic material itself as previously assumed, but is also strongly influenced by the interface between the metal capping layer and the ferromagnets opening new paths to engineering the SSE signals.

**Acknowledgements**

The authors would like to express their great thanks to Dr. Alexander Serga and Prof. Burkard Hillebrands in TU Kaiserslautern for providing a portion of LPE YIG samples and to Prof. Gerrit E. W. Bauer and Dr. Sebastian T. B. Goennenwein for valuable discussions. Also, we want to thank the financial support through Deutsche Forschungsgemeinschaft (DFG) SPP 1538 "Spin Caloric Transport", the Graduate School of Excellence Materials Science in Mainz (MAINZ) and the EU projects (IFOX, NMP3-LA-2012246102, INSPIN FP7-ICT-2013-X 612759, MASPIC ERC-2007-StG 208162).

**Author contributions**

E. G. and A. K. fabricated the devices, performed the measurements and analyzed the data. J. C. assisted in the sample preparation. G. J. contributed to the magnetron sputtering and the experimental setups. G. J. and M. K. planned and supervised study. E. G., A. K., and M. K. wrote the manuscript. All authors discussed the results and commented on the manuscript.

**Competing financial interests**

The authors declare no competing financial interests.

**Figure Captions**

**Figure 1.** (color online) **Temperature dependent SSE coefficients ($\sigma_{SSE}$) in YIG.** (a) for a YIG slab and (b) for YIG thin films with a wide range of thicknesses. The open symbols in (a) represent the results recorded by field sweep mode measurements. (c) The peak position of the $\sigma_{SSE}$-$T$ curves as a function of film thickness. The inset shows the schematic diagram of the sample structure and measurement setup.

**Figure 2.** (color online) **The calculated thermally excited magnon mean propagation length.** Thickness dependent $\sigma_{SSE}$ measured at a fixed temperature of (a) 250 K, (b) 120 K, and (c) 50 K, respectively. The solid lines are the fitting results according to the atomistic spin model. (d) Temperature dependent magnon mean propagation length $\xi$. The solid line is the fit to $\xi \sim T^{-1}$.

**Figure 3.** (color online) **Magnetic field and heating power dependence of the SSE.** (a) Magnetic field dependent normalized SSE signals of Pt/Cu/YIG hybrids at various temperatures from 30 K to room temperature. (b) Temperature dependent field suppression ratio of Pt/Cu/YIG hybrids. (c) Temperature dependent $\sigma_{SSE}$ of Pt/Cu/YIG hybrids measured at a fixed heating power and varied magnetic fields from 0.2 T to 9 T. (d) Temperature dependent SSE voltages of Pt/Cu/YIG hybrids with a fixed magnetic field of ± 0.2 T and different heating currents varied from 5 to 9 mA.

**Figure 4.** (color online) **Temperature dependence of the SSE with different capping layers.** Temperature dependent $\sigma_{SSE}$ measured at Pt/YIG, Pt/Cu/YIG, and Pd/YIG hybrids. The thickness of YIG films is 50 μm. The inset shows the normalized value of $\sigma_{SSE}/\sigma_{SSE,Max}$ to the maximum values at the SSE peak temperatures.

**FIG. 1.**

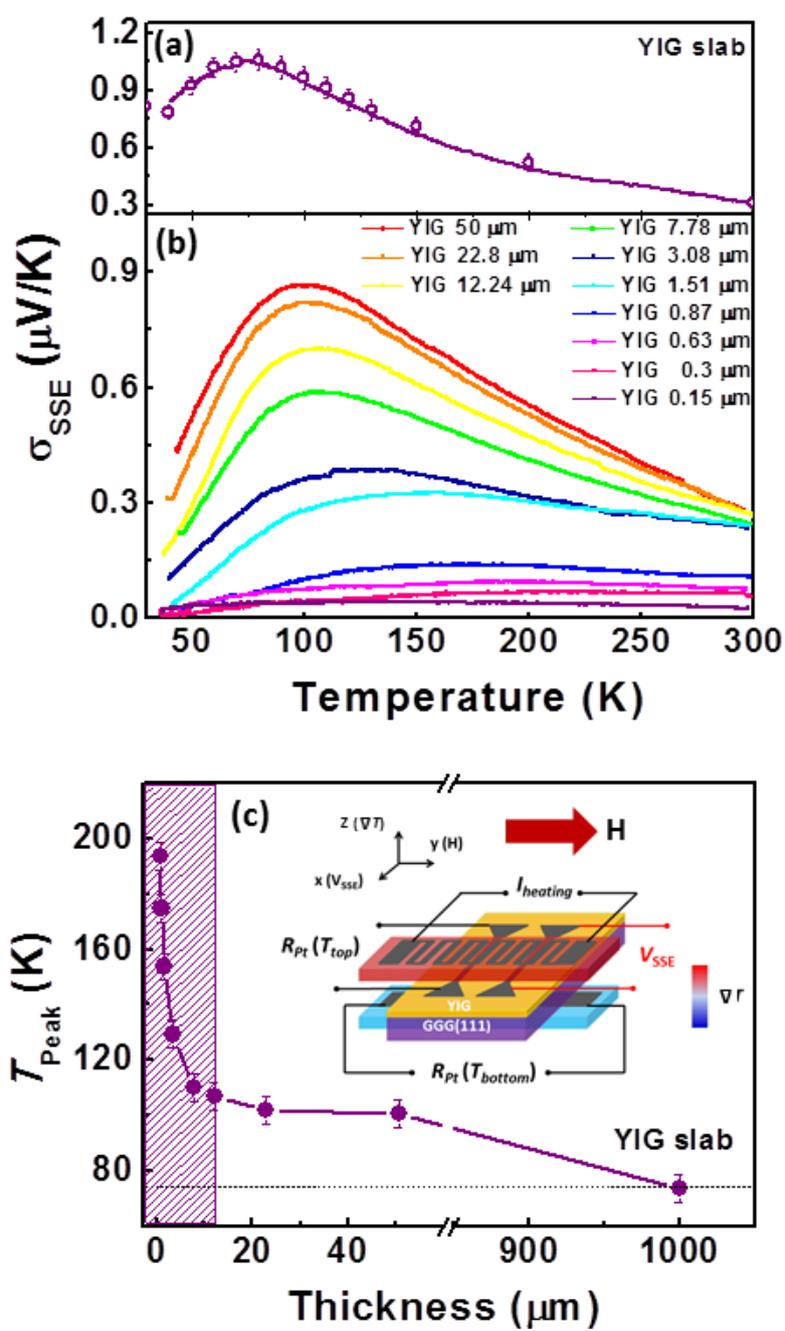

**FIG. 2.**

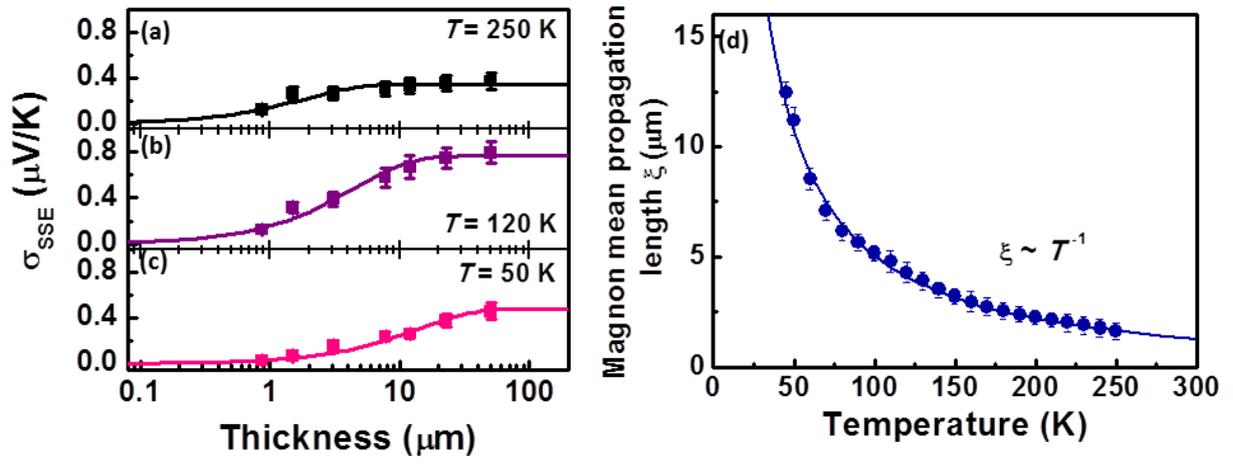

**FIG. 3.**

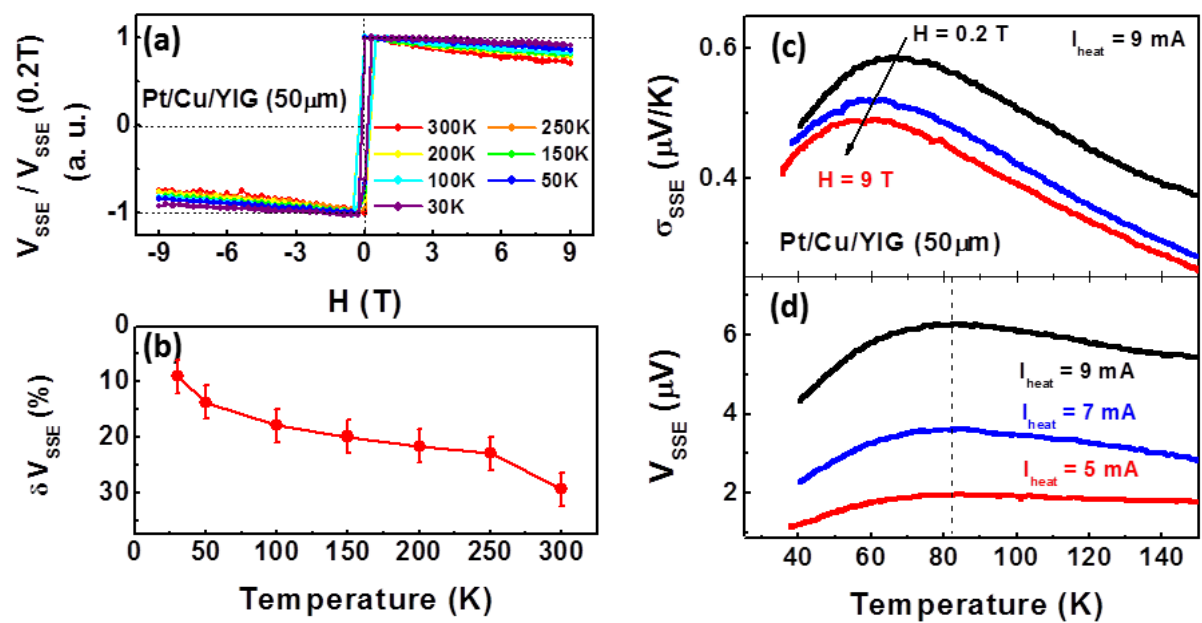

**FIG. 4.**

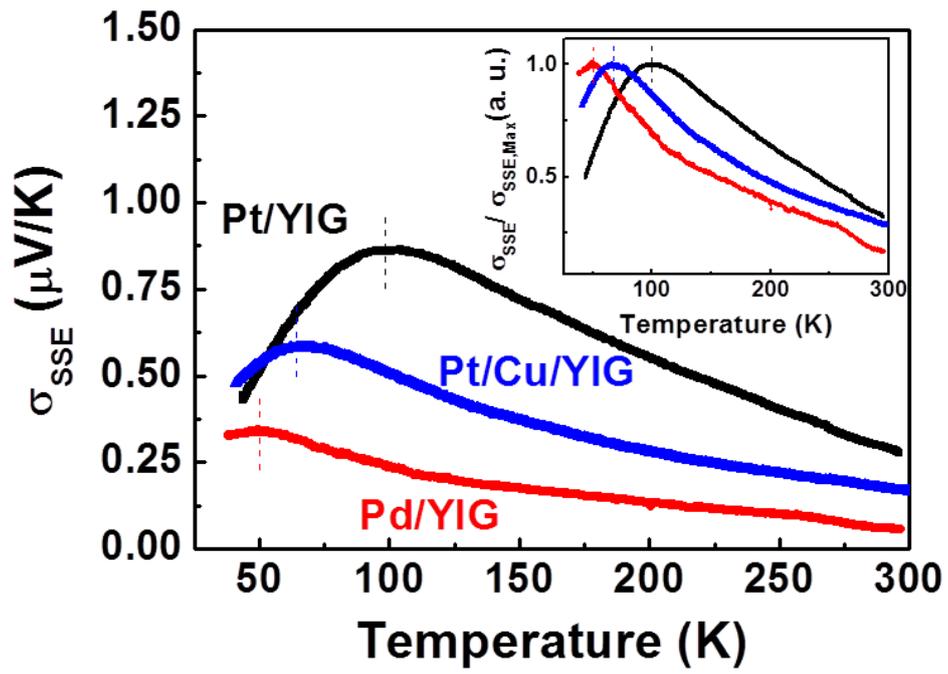